# Plasticity, localization, and damage in ferritic-pearlitic steel studied by nanoscale digital image correlation

## T. Vermeij[1] & J.P.M. Hoefnagels[1,*]

*[1] Dept. of Mechanical Engineering, Eindhoven University of Technology, The Netherlands*

*[*]Corresponding author: j.p.m.hoefnagels@tue.nl*

## Abstract

The evolution of deformation from plasticity to localization to damage is investigated in ferritic-pearlitic steel through nanometer-resolution microstructure-correlated SEM-DIC (μ-DIC) strain mapping, enabled through highly accurate microstructure-to-strain alignment. We reveal the key plasticity mechanisms in ferrite and pearlite as well as their evolution into localization and damage and their relation to the microstructural arrangement. Notably, two contrasting mechanisms were identified that control whether damage initiation in pearlite occurs and, through connection of localization hotspots in ferrite grains, potentially results in macroscale fracture: (i) cracking of pearlite bridges with relatively clean lamellar structure by brittle fracture of cementite lamellae due to build-up of strain concentrations in nearby ferrite, versus (ii) large plasticity without damage in pearlite bridges with a more "open", chaotic pearlite morphology, which enables plastic percolation paths in the interlamellar ferrite channels. Based on these insights, recommendations for damage resistant ferritic-pearlitic steels are proposed.

## Graphical Abstract

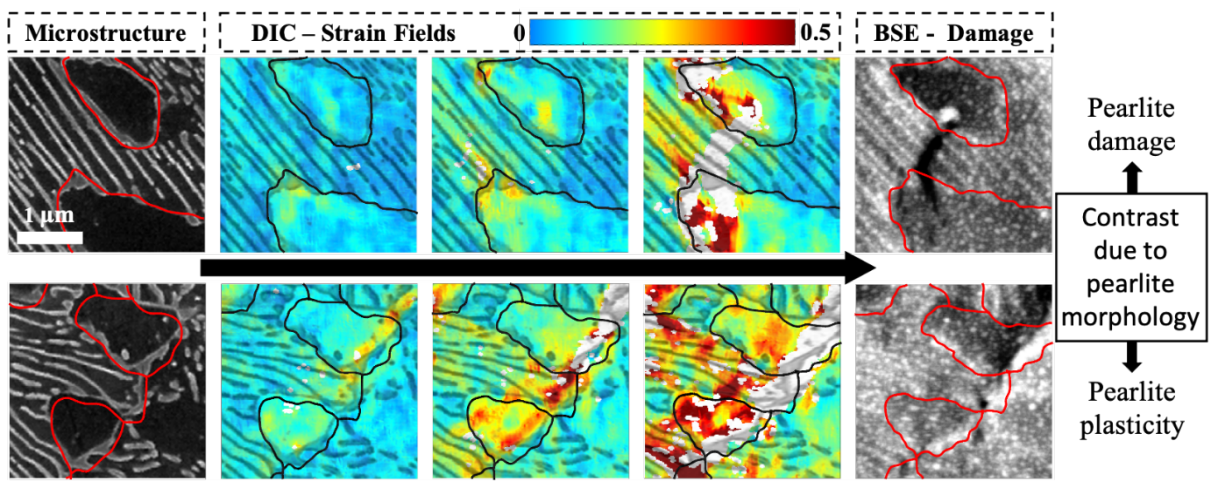

**Keywords:** Ferritic-pearlitic steel, nanoscale digital image correlation, plasticity, damage, localization

Ferritic-pearlitic steels are strong and formable through the combined effects of ferrite grains and pearlite colonies, comprising alternating lamellae of randomly distributed and oriented cementite within a ferritic matrix [1,2]. Deformations in ferritic-pearlitic steels intrinsically arise from the nanometer scale arrangements of cementite and ferrite, while the ferritic-pearlitic distribution determines how plasticity and damage will propagate and coalesce, ultimately leading to failure. In the literature, challenges have arisen from the intrinsic multi-scale nature of deformations in ferritic-pearlitic steel, inhibiting identification of deformation mechanisms and potential experimental-numerical comparisons [3–10].

Deformations in ferritic-pearlitic steels can be observed through *in-situ* mechanical testing in a scanning electron microscope (SEM) [3,6–8]. For example, pearlite plasticity was found to be controlled by ferrite deformation, with damage occurring through cementite cracking, likely caused by dislocation pile-up at the ferrite-cementite interfaces [3,7]. Previously, we employed digital image correlation (DIC) on SEM micrographs to measure strain fields in banded microstructures in dual-phase (ferrite-martensite) and ferritic-pearlitic steels, wherein pearlite appeared to deform plastically [6]. However, the SEM microstructure maps that were employed as DIC patterns limited the spatial resolution of the strain fields to several micrometers, which was insufficient to resolve deformations at lamellar scales. Such studies (and others) reveal limited and contrasting conclusions on the role of cementite lamellae on pearlite deformation, moreover, the full evolution (or lack thereof) of plasticity to damage is rarely studied.

More recently, advances in DIC speckle patterning techniques enables SEM-DIC ($\mu$-DIC) for highly robust measurements of localized strains at micrometer and even nanometer scales [11–14]. Such high spatial strain resolutions permit experimentalists to resolve plasticity between (or inside) cementite lamellae and provides the opportunity to simultaneously monitor damage initiation in the microstructure [12,13]. Damage and failure mechanisms at the micro- and nanoscale are notoriously complex and can only be unraveled by relating them to the preceding unique plastic deformation mechanisms, in order to provide proper understanding to enable accurate model prediction and/or material design optimization. Plasticity and damage have been studied for the same material (e.g. [9]), however, not both at the same location in the microstructure. In fact, to our knowledge, the complete full-field evolution from preceding microscale plasticity to resulting damage has yet to be studied for ferritic-pearlitic steels.

Here, we measure strains with high-resolution and robust SEM-DIC all the way up to damage and local fracture. Careful alignment between ferrite-pearlite-cementite morphology and nanometer resolution strain fields reveals how plasticity localizes in relation to ferritic-pearlitic interfaces and (individual) cementite-ferrite lamellae. Upon continued deformation, damage occurs at critical microstructure locations. These are unraveled on the basis of the preceding strain evolution, providing insights in damage mechanisms. Finally, we discuss potential strategies for increased robustness to early damage and failure in ferritic-pearlitic steels.

A ferritic-pearlitic steel plate (0.17C-0.16Si-1.27Mn-0.014P-0.012Cr-0.037Al-0.038Ni wt.%) was retrieved after 45 minutes austenization at 900 °C and subsequent cooling in the furnace. Samples were cut and metallographically prepared to a final mechanical polish (OPS colloidal silica particles), followed by a 5-seconds, 2%-Nital etching step. Mapping of the microstructural arrangement of ferrite, pearlite and cementite was performed in a Tescan Mira 3 SEM using backscattered electron (BSE) imaging. Figure 1(a) shows the chosen region of interest with ~20 $\mu$m ferrite grains and pearlite with ~225 nm lamellar spacing. The crystallographic orientation of all ferrite phase was mapped with EBSD (Edax Digiview 2 camera) at 50 nm pixelsize, in order to identify the ferrite grain orientations and grain boundaries as well as the orientation of the ferrite matrix within the pearlite. Subsequently, to enable high-resolution SEM-DIC strain measurements at lamellar scales, a well-controlled, highly dense, nanometer sized speckle pattern was applied using a recently proposed novel patterning method, i.e. single-step sputter deposition of a low temperature solder alloy that naturally forms nanoscale islands during deposition [13]. Specifically, In52Sn48 was sputtered at 15 mA, 1E-2 mbar, 25 °C, 180 s, according to (c) in Table 1 of Hoefnagels *et al.* [13], resulting in a random, dense, high-quality pattern with features ~25-100 nm in size as shown in Figure 1(b). Thereafter, the sample was deformed through uniaxial tension using a micro-tensile stage (Kammrath & Weiss) in the SEM. To capture the strain fields, the DIC pattern was imaged intermittently using 5 kV in the region of interest at 15 nm pixel size, according to guidelines given in [13]. SEM images of 4096 by 4096 pixels resulted in a 60×60 μm² field of view.

DIC was performed with VIC-2D software using the following main parameters: subset size of 21 pixels (325 nm), step size 1 pixel (15 nm) and linear subset shape functions, see Hoefnagels *et al.* [13] and the DIC guide [15] for more details. This small step size appears to yield the highest possible spatial strain resolution, which is critical to distinguish the high strain levels in the ferrite lamellae in pearlite from the low strain levels in the adjacent cementite lamellae. The Green-Lagrange strain tensor $\boldsymbol{\varepsilon} = \frac{1}{2}\left[\left(\overrightarrow{\nabla_0 u}\right)^T + \overrightarrow{\nabla_0}\vec{u} + \left(\overrightarrow{\nabla_0}\vec{u}\right)^T \cdot \overrightarrow{\nabla_0}\vec{u}\right]$ is calculated from the DIC displacement field, $\vec{u}(\vec{x})$, through the (numerically computed) displacement gradient tensor, $\nabla_x \vec{u}$. We use the 2D equivalent von Mises strain measure, defined as $\varepsilon_{vM}(\vec{x}) = \frac{\sqrt{2}}{3}\sqrt{\left(\varepsilon_{xx} - \varepsilon_{yy}\right)^2 + \varepsilon_{xx}^2 + \varepsilon_{yy}^2 + 6\varepsilon_{xy}^2}$, as it is a good indicator for (local) plasticity [16]. Highly accurate alignment of the strain fields to the ferritic-pearlitic microstructural morphology is required to attribute the domains of high/low strains to, respectively, large/small plasticity in the ferrite/cementite lamellae. Therefore, micrographs were captured at 20 kV (before tensile testing) to image both the DIC speckle pattern and the underlying ferritic-pearlitic morphology (not shown) for the microstructure to strain fields alignment. Linear interpolation was employed through several selected homologous points in both microstructure datasets (before and after pattern application), yielding a "distortion" field with which the full dataset was aligned. In contrast to SEM images, the EBSD map contains severe spatial distortions [17], therefore, a novel procedure based on global-DIC [18] was used to align the EBSD map to the microstructure, which enables to plot the ferrite grain boundaries as overlay to all figures in this work. To intuitively link strains to the ferritic-pearlitic morphology, we plot the aligned microstructure maps over the strain fields as an overlay with variable transparency, such that only the cementite lamellae are clearly visible (using a Jet colormap that is cut off to distinguish the cementite), see Figure 1(c). Note that all presented strain fields are plotted in the deformed configuration, by forward-deforming the microstructure and strain field datasets (from their reference configuration) using the DIC displacement field.

To enable nanoscale damage analysis, (post-mortem) BSE scans were acquired at 20 kV, after conclusion of the *in-situ* test, such that the electron beam protrudes the nanometer thickness InSn DIC pattern. The post-mortem 20 kV BSE scan is aligned to the (last increment of the) forward-deformed 5kV SEM scan of the microstructure, by selection of homologous points (such as clearly recognizable undamaged cementite lamellae). After all processing steps, the aligned data was imported in MTEX [19] for consistent plotting requiring no manual adjustments, such as cropping, alignment, etc.

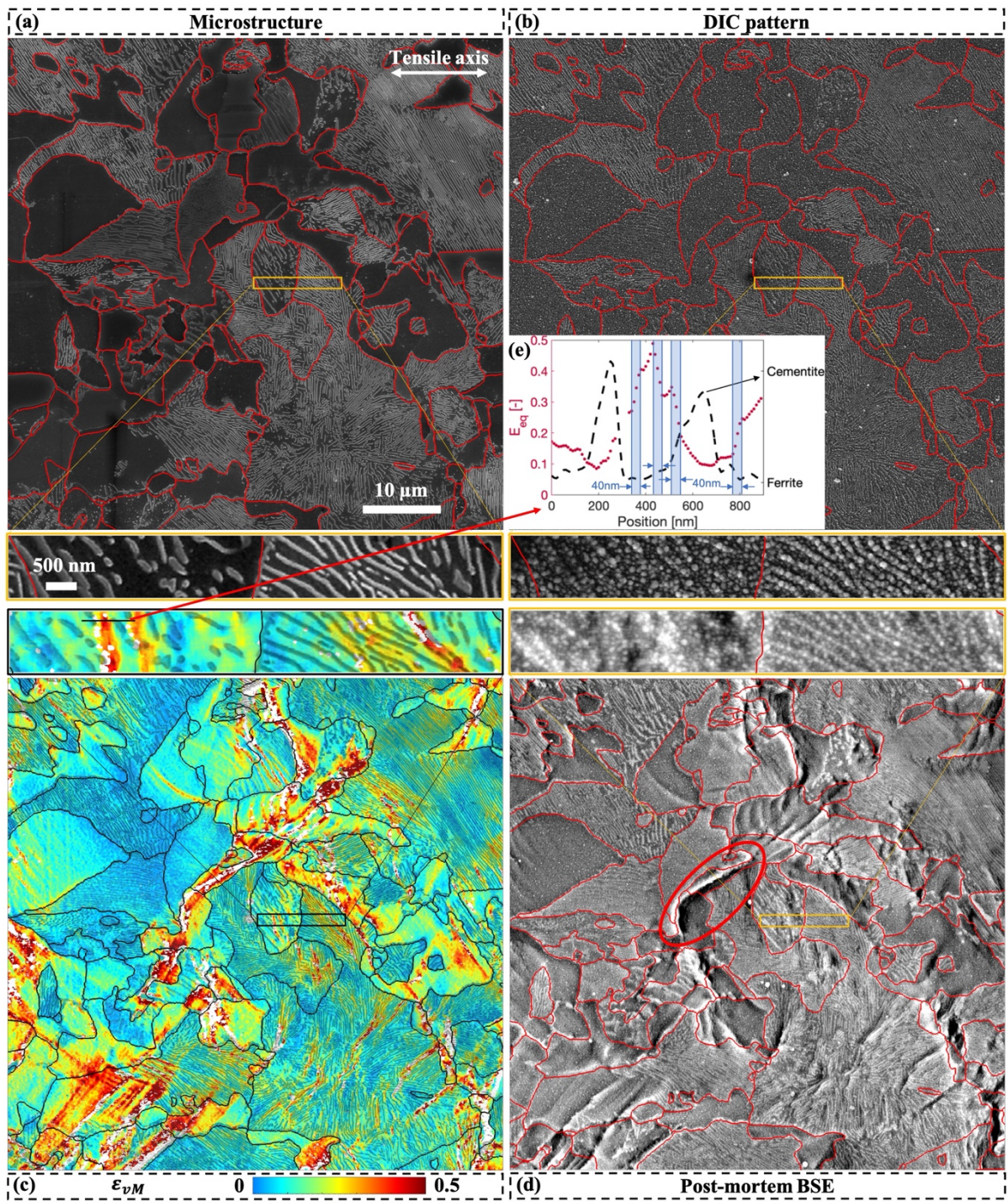

*Figure 1: Overview of microstructure and deformations of the complete region of interest: (a) BSE microstructure map, showing ferrite grains, pearlite colonies and cementite lamellae. (b) DIC pattern on the same region of interest. (c) SEM-DIC equivalent strain field at a global von Mises strain of $\overline{\varepsilon_{vM}} = 0.19$. The ferritic-pearlitic morphology is mapped as a transparent overlay, with stronger transparency for ferrite than cementite and grayscale inverted. (d) Post-mortem BSE image shows damage (at $\overline{\varepsilon_{vM}} = 0.36$), with a pronounced damage event circled in red. For each dataset, a magnified inset of the same area is included to highlight the alignment, spatial resolution and quality of the data; note also the small scalebar of 500 nm. Grain boundaries of ferrite, derived from a spatial-distortion corrected EBSD scan, have been overlayed as red (a,b,d) and black (c) lines. In insert (e) strain and microstructure data is plotted over a line profile taken from the magnified inset of (c), showing nanoscale strain partitioning between cementite and ferrite lamellae.*

The evolving equivalent ($\varepsilon_{vM}$) strain fields are computed for 4 subsequent global deformation levels of: $\overline{\varepsilon_{vM}} = 0.07$, $\overline{\varepsilon_{vM}} = 0.12$, $\overline{\varepsilon_{vM}} = 0.19$ and $\overline{\varepsilon_{vM}} = 0.36$. The global *von Mises* strain $\overline{\varepsilon_{vM}}$ is defined as the 2D equivalent von Mises strains averaged over the FOV. Therefore, we do not rely on the nominal strain, derived from the clamp displacements, as it is unreliable due to the heterogeneous deformations over the gauge section and slip in the clamps. The global engineering stresses at these deformation steps were 250 MPa, 475 MPa, 525 MPa and 550 MPa, respectively. The strain field at $\overline{\varepsilon_{vM}} = 0.19$ in Figure 1(c) shows how the ferritic-pearlitic microstructure deforms highly heterogeneously. As expected, strains proceed through the microstructure under angles of approximately ±45° with respect to the horizontal loading direction, predominantly propagating through ferrite channels between pearlite colonies. Indeed, ferritic-pearlitic strain partitioning is the dominant deformation mechanism which appears to evolve into ferritic-pearlitic interface "decohesion", as observed in the center of Figure 1(d), indicated with the red ellipse. However, detailed analysis of the BSE contrast, which shows darker regions in the ferrite grains near the pearlite colonies in Figure 4($b_5$) and ($c_5$), indicates these to be so-called "distributed nano-damage" (similar to high-resolution observations of ferritic-martensitic steels, identifying distributed nano-damage in ferrite near martensite as the main damage mechanism [20]). For both types of steel, however, interface "decohesion" between ferrite and the 'hard' phase (pearlite or martensite) is typically reported as the main damage mechanism, which is likely caused by the scales of observation as distributed nano-damage can only be recognized at high magnification.

Next, we focus on the high spatial resolution of the strain fields. Particularly, strain partitioning in regions with interlamellar spacing <100 nm shows that we can resolve strains at these small dimensions. To clarify the actual spatial strain resolution and the effectiveness microstructure-to-strain alignment, Figure 1(e) shows a line plot of microstructure and strain data taken over the line shown in the inset of Figure 1(c). The dashed black line shows the SEM image intensity, corresponding to cementite at high values and ferrite at low values. The equivalent strain data is shown through the dotted red line, with several steps having a width of ~40 nm as marked by the blue boxes, giving an indication of the resolution [21]. Furthermore, the effectiveness of microstructure-to-strain alignment is validated here by the occurrence of strain peaks between cementite lamellae, as expected. Even though the spatial strain resolution appears to be as low as ~40 nm, one should be careful in assessing the strain magnitude in the lamellae for lamellar spacings of similar magnitude as the strain resolution, which is not always taken into account in literature.

Next, we investigate how plasticity propagates through pearlite colonies. Typically, plasticity initializes in ferrite as hotspots near pearlite. Upon increased loading, these hotspots need to be connected, often requiring percolation through pearlite colonies. In Figure 2, two smaller areas (taken from the larger region of interest in Figure 1) show two contrasting pearlite deformation modes. For percolation paths aligned with the pearlite lamellar direction, see Figure 2(a), the plasticity hotspots can connect by protruding between the cementite lamellae, through well aligned or chaotic ("open") lamellae structures. The deformation is localized predominantly in interlamellar ferrite, only occasionally crossing single cementite lamellae, resulting in a strong cementite-ferrite strain partitioning. In contrast, Figure 2(b) highlights examples of deformation percolation paths perpendicular to the pearlite lamellar direction, where the deformation is inhibited by continuous, non-chaotic cementite lamellae. In this case, as the cementite lamellae block the formation of localization bands inside the interlamellar ferrite, the deformation is smeared out over wide, intermittent, bands with much lower cementite-ferrite partitioning. Analysis of the principal strain fields (not shown) reveals that these bands deform approximately by the globally applied deformation, which is uniaxial tension, in contrast to the percolation paths in Figure 2(a). In essence, the clear contrast between these two configurations can be loosely interpreted as the difference in deformation of a parallel chain (Figure 2(a)) versus a series chain (Figure 2(b)) under tension.

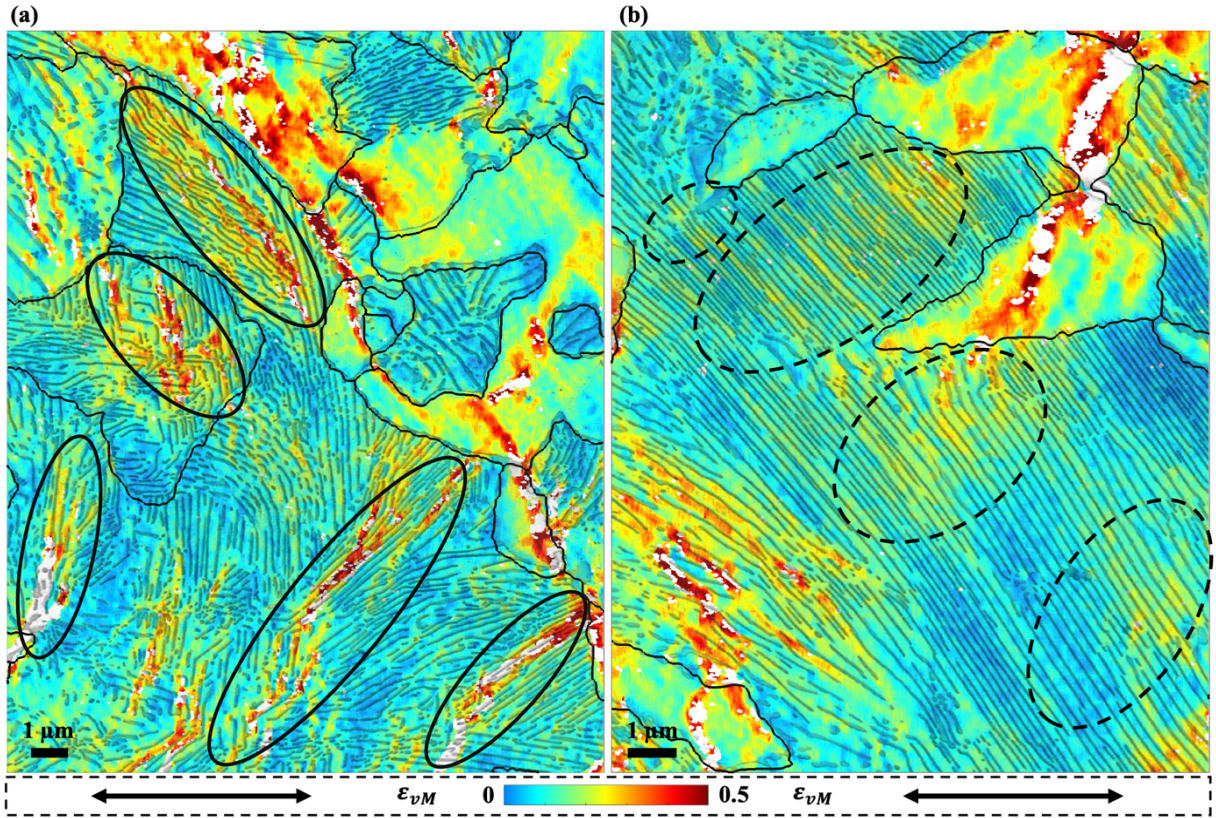

*Figure 2: Equivalent strain maps (with transparent microstructure overlay) showing progression of plasticity through pearlite at $\overline{\varepsilon_{vM}} = 0.19$, for two contrasting cases: (a) pearlite percolation bands along the lamellar direction, showing clear evidence of cementite-ferrite strain partitioning with sharp, continuous plasticity bands localized between the cementite lamellae, as highlighted by the solid ellipses, and (b) pearlite percolation bands perpendicular to the lamellar direction, showing smeared out, interrupted plasticity bands in pearlite, as highlighted by the dashed ellipses. Ferrite grain boundaries from EBSD are overlayed as black lines.*

Upon further deformation, damage events are observed in the microstructure. We focus on a selection of pearlite "bridges" (i.e. narrow pearlite areas surrounded by ferrite) that show significant plasticity and damage, as these are prone to act as percolation pathways between deformation hotspots in nearby ferrite grains. When these pearlite bridges undergo (early) local failure, it enables meso-scale localization bands to form, that are likely responsible for early damage propagation leading up to global failure. Two distinct deformation modes are identified: high plasticity without pearlite cracking (Figure 3) and clear pearlite cracking/fracture events (Figure 4). The nanometer level strain resolution enables detailed analysis of the evolution of strain localization and/or strain partitioning that does, or does not, lead to damage. In both figures, subfigure ($x_1$) shows the microstructure, subfigures ($x_2$, $x_3$, $x_4$) show the SEM-DIC equivalent strain fields, and subfigure ($x_4$) shows damage through post-mortem BSE micrographs.

Figure 3 shows three examples of pearlite bridges that deform significantly without showing clear fracture. Indeed, in Figure 3(a) and (b) there are no signs of any damage in the BSE micrographs, while plasticity proceeds mostly through the interlamellar ferrite, allowed by the "open" pearlite arrangement. In Figure 3(c), the localization crosses the pearlite bridge through the vertical interlamellar ferrite channel, by overcoming the location where two cementite lamellae touch (indicated by the black arrow). These touching lamellae move apart, causing localized plasticity (Figure 3($c_3$) and ($c_3$)), ultimately causing local damage (Figure 3($c_5$)), without however resulting in cracking of the full pearlite bridge. The plastic localization across pearlite bridges (inhibition of damage) seems to be uniquely tied to the two-phase (ferritic-pearlitic) microstructure, e.g. in fully pearlitic steel, plastic localization across pearlite colonies does not seem to be relevant [3].

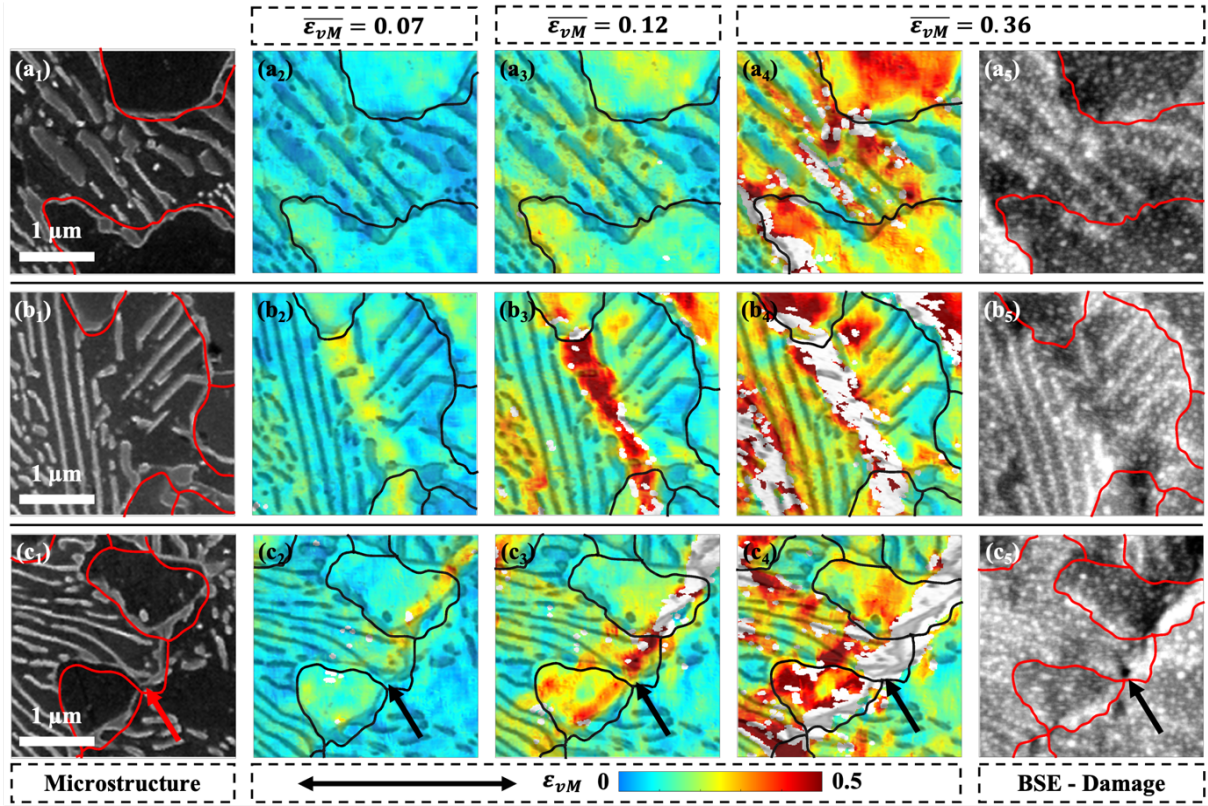

*Figure 3: Three examples of pearlite "bridges" that do not show clear pearlite cracking, investigated at high spatial resolutions (a-b-c). ($a_1b_1c_1$) The microstructure with bright cementite lamellae, with the ferrite grain boundaries from EBSD overlayed as red lines; SEM-DIC equivalent strain fields (with transparent overlaid microstructure) at ($a_2b_2c_2$) $\overline{\varepsilon_{vM}} = 0.07$, ($a_3b_3c_3$) $\overline{\varepsilon_{vM}} = 0.12$ and ($a_4b_4c_4$) $\overline{\varepsilon_{vM}} = 0.36$; ($a_5b_5c_5$) post-mortem BSE image, showing damage at $\overline{\varepsilon_{vM}} = 0.36$. The arrows in (c) mark a location where two touching cementite lamellae move apart. The ferrite grain boundaries (black or red lines) are not perfectly aligned and are for illustrative purposes only.*

In contrast, Figure 4 shows three examples of clear fracture of pearlite bridges with apparently more regular and cleaner lamellar configurations than those in Figure 3. In Figure 4(a) and (b), plasticity evolves in ferrite, at two sides of the eventually cracking pearlite as indicated by arrows, while the strains inside the pearlite remain rather low as a few parallel cementite lamellae block the localization band. Thus, this pearlite cracking mechanism appears to originate from (rather) brittle fracture of multiple cementite lamellae, forced by increasing levels of surrounding (ferrite) deformation, which agrees well with observations in coarse fully pearlitic steel [3]. Figure 4(c) shows similar behavior, although the localization band first propagates through an interlamellar ferrite channel in the pearlite bridge that is blocked by 2-3 remaining cementite lamellae (Figure 4($c_3$)), which again abruptly crack in a collective manner at the final deformation stage. These observations reveal that the interaction between ferrite and pearlitic is critical, resulting in key microscale deformation and damage mechanisms in ferritic-pearlitic steel that do not occur in fully ferritic or fully pearlitic microstructures. Deformation may initialize in ferrite grains and channels, but will inevitably interact with pearlite colonies, activating various deformation mechanisms and ultimately leading to damage and failure.

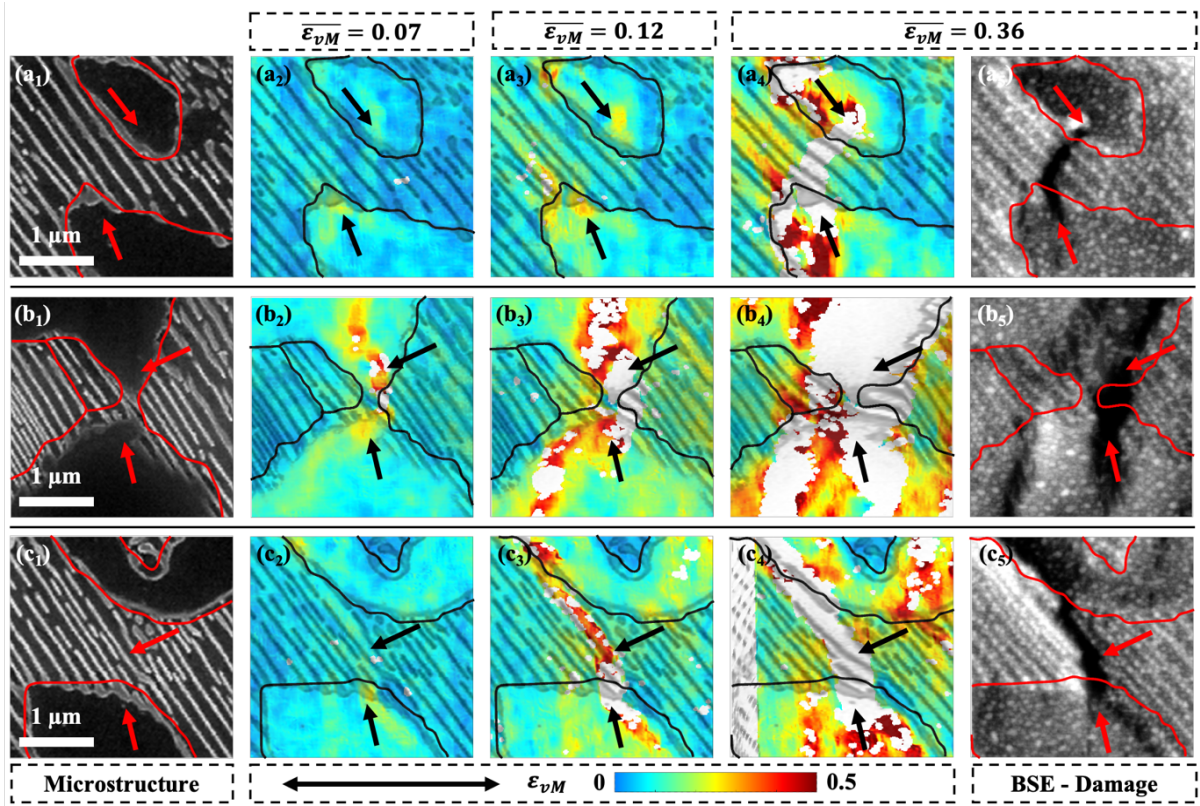

*Figure 4: Three examples of pearlite "bridges" that clearly show pearlite cracking, investigated at high spatial resolutions (a-b-c). ($a_1b_1c_1$) The microstructure with bright cementite lamellae, with the ferrite grain boundaries from EBSD overlayed as red lines; SEM-DIC equivalent strain fields (with transparent overlaid microstructure) at ($a_2b_2c_2$) $\overline{\varepsilon_{vM}} = 0.07$, ($a_3b_3c_3$) $\overline{\varepsilon_{vM}} = 0.12$ and ($a_4b_4c_4$) $\overline{\varepsilon_{vM}} = 0.36$; ($a_5b_5c_5$) post-mortem BSE image, showing damage at $\overline{\varepsilon_{vM}} = 0.36$. The arrows mark locations of strain localization outside of a (partial) pearlite bridge, resulting in catastrophic pearlite cracking. The ferrite grain boundaries (black or red lines) are not perfectly aligned and are for illustrative purposes only.*

Finally, we discuss potential improvements for the design of future ferritic-pearlitic steels. Pearlite bridges delay easy percolation of plasticity through the compliant ferrite grains and thereby provide increased strength to the material, however, subsequent cracking of these pearlite bridges needs to be suppressed to increase global ductility. Therefore, we recommend tailoring of the microstructure morphology by tuning the processing conditions.

Cementite lamellae in pearlite bridges should be configured to allow percolation of localized plasticity, e.g. through an "open" pearlite arrangement as in Figure 3(a) and (b) (could be attempted by slower furnace cooling) or by interlamellar ferrite channels that cross the pearlite bridges as in Figure 3(c). Alternatively, the pearlite bridges can be designed to be longer (for example by increasing ferrite grain size) to significantly increase the average number of bridge-crossing percolation paths. Note, however, that the width of these paths should remain small enough to enforce localized ferrite plasticity and hardening, in order to preserve the global strength. Additionally, refinement of the pearlite has been reported to improve ductility [3] and could therefore also be worth exploring. Interestingly, these recommendations align with the suggestions for damage inhibition strategies in other multi-phase microstructures (ferritic-martensitic [22,23], martensitic-austenitic [24] and Ti-Al-V-Fe [25] alloys), wherein the ideal response of the 'hard' phase combines a high yield strength (thus high instead of low phase contrast) with an internal plastic mechanism that can prolong plasticity.

In summary, we successfully employed nanometer-resolution microstructure-correlated SEM-DIC strain mapping of the full evolution of strain localization to damage at lamellar scales in ferritic-pearlitic steel. This high-quality strain quantification could only be achieved with: (i) a high-quality SEM-DIC speckle pattern yielding accurate determination of the high levels of plastic deformation occurring just before and around damage events, (ii) successful high-eV BSE characterization of damage without hindrance from the DIC pattern, and (iii) careful alignment of the sub-micrometer microstructure morphology, strain fields, and damage maps in the deformed configuration.

Using this novel approach, a variety of deformation micro-mechanisms was observed. Notably, two contrasting mechanisms were identified that control whether damage initiation in pearlite occurs and, through connection of localization hotspots in ferrite grains, potentially results in macroscale fracture: (i) cracking of pearlite bridges with relatively clean lamellar structure by brittle fracture of cementite lamellae due to build-up of strain concentrations in nearby ferrite, versus (ii) large plasticity without damage in pearlite bridges with a more "open", chaotic pearlite morphology, which enables plastic percolation paths in the interlamellar ferrite channels. Based on these insights, recommendations for damage resistant ferritic-pearlitic steels are proposed.

## Acknowledgements

The authors thank Marc van Maris for experimental support. This research was carried out under project number S17012b in the framework of the Partnership Program of the Materials innovation institute M2i (www.m2i.nl) and the Technology Foundation TTW (www.stw.nl), which is part of the Netherlands Organization for Scientific Research (http://www.nwo.nl).